\documentclass[useAMS,usenatbib,usegraphicx]{mn2e}
\usepackage{multirow}
\title[Photometry, kinematics and stellar 
populations in NGC\,357]{Constraining the formation of inner bars.\\
Photometry, kinematics and stellar 
populations in NGC\,357\thanks{Based on observations carried out at the European Southern
Observatory (ESO 70.B-0338).}.}
\author[de Lorenzo-C\'aceres et al.]
{A. de Lorenzo-C\'aceres$^{1,2}$\thanks{adlcr@iac.es}, A. Vazdekis$^{1,2}$\thanks{vazdekis@iac.es}, 
J.~A.~L. Aguerri$^{1,2}$\thanks{jalfonso@iac.es},\and
E.~M. Corsini$^{3}$\thanks{enricomaria.corsini@unipd.it} and Victor P. Debattista$^{4}$\thanks{vpdebattista@gmail.com}\\
$^{1}$Instituto de Astrof\'isica de Canarias (IAC), E-38205 La Laguna, Tenerife, Spain\\
$^{2}$Depto. Astrof\'isica, Universidad de La Laguna (ULL), E-38206 La Laguna, Tenerife, Spain\\
$^{3}$Dipartimento di Astronomia, Universit\`a di Padova, vicolo dell'Osservatorio 2, I-35122 Padova, Italy\\
$^{4}$RCUK Fellow, Jeremiah Horrocks Institute, University of Central Lancashire, Preston PR1 2HE, UK}

\begin{document}

\date{Accepted xxx. Received xxx; in original form xxx}

\pagerange{\pageref{firstpage}--\pageref{lastpage}} \pubyear{2002}

\maketitle

\label{firstpage}

\begin{abstract}
Double-barred galaxies are common in the local Universe, with approximately one third
of barred spirals hosting an smaller, inner bar. Nested bars have been proposed as a mechanism
for transporting gas to the very central regions of the galaxy, trigger star formation and contribute to the
growth of the bulge. To test this idea, we perform for the first time a detailed analysis
of the photometry, kinematics and stellar populations of a double-barred galaxy: NGC\,357.
We find that this galaxy is either hosting a pseudobulge
or a classical bulge together with an inner disc.
We compare the relative mean luminosity-weighted age, metallicity
and $\alpha$-enhancement between the (pseudo)bulge, inner bar and outer bar, finding
that the three structures are nearly coeval and old. Moreover, the 
bulge and inner bar present the same metallicity and overabundance, whereas the outer bar
tends to be less metal-rich and more $\alpha$-enhanced. 
These results point out that, rather than the classical secular scenario in which 
gas and star formation play a major role, the redistribution of the 
existing stars is driving the formation of the inner structures.
\end{abstract}

\begin{keywords}
galaxies: individual: NGC\,357 --- galaxies: photometry --- galaxies: kinematics and dynamics --- 
galaxies: bulges --- galaxies: structure --- galaxies: evolution.
\end{keywords}

\section{Introduction}

The structural evolution of disc galaxies and its timescale remain as open questions
nowadays. In fact, it is still not
clear if the formation of bulges is mostly driven by external processes, such as interactions or mergers, or 
by the internal secular evolution due to inflow of gas from the galactic disc to the central regions,
where it may trigger star formation 
\citep[see][for a review]{2004ARA&A..42..603K}. This second
scenario implies a slow process that
takes place when non-axysimmetric components (particularly bars) redistribute the angular
momentum of the galaxy. Given the large number of barred
galaxies found in the local Universe 
\citep{2000ApJ...529...93K,2000AJ....119..536E,2007ApJ...659.1176M,2007ApJ...657..790M,2009A&A...495..491A}, 
and that the bar fraction remains significant till z$\sim$1
\citep[although its evolution with redshift is still a matter of debate, see e.g.][]
{2004ApJ...612..191E,2004ApJ...615L.105J,2008ApJ...675.1141S,2010MNRAS.409..346C}, 
it seems that internal secular processes might actually be an important driver of the evolution in spirals.
If this is the case, the resulting kinematical and morphological changes 
could also lead to an evolution along the Hubble sequence from later to earlier types, as 
suggested by \citet{1993A&A...277...27F}.

It has been theoretically demonstrated that the flow of gas through a single bar
stops before reaching the galactic centre; however, a system of nested bars
may efficiently transport the material to the nuclear regions, where it even would
help to fuel nuclear activity. This idea was first proposed by 
\citet{1989Natur.338...45S,1990Natur.345..679S}, and resulted in an increasing interest in
galaxies hosting two or more bars.

Double-barred galaxies are common objects in the local Universe; at least one third of 
barred galaxies show an additional inner bar 
\citep{2001BSAO...51..140M,2002AJ....124...65E,2002ApJ...567...97L,2004A&A...415..941E}
and nested bars have been found till \emph{z}$\sim$0.15 \citep{2006MNRAS.370..477L}.
Photometric studies have shown that the two bars are randomly oriented relative to each other
\citep{1993A&A...277...27F,1995A&AS..111..115W}, suggesting that they rotate independently, as
predicted by numerical simulations \citep{2002ApJ...565..921S,2007ApJ...654L.127D}
and confirmed by observations \citep{2003ApJ...599L..29C}. 
Moreover, it seems there is no correlation between the presence of an inner bar and the 
Hubble type of the host galaxy or its main bar properties. 
\citet{2002AJ....124...65E} find that inner bars extend to $\sim$12\% of the outer bar size,
and that have in general smaller ellipticities, but may be larger since the 
inner bar might be embedded in the light of the bulge.

\citet{1997ApJ...484L.117M,2000MNRAS.313..745M} study the dynamics of double-barred objects
and develop a simple formalism to study the orbits of their stars based on the \emph{loop} concept.
Single bars are characterized by two fundamental frequencies (one for the bar and one for the 
free oscillations), allowing the existence
of closed and periodic orbits. In fact, two orbital families stand out for a single-barred 
potential: the \emph{x$_1$} orbits, elongated along the bar major axis, and the \emph{x$_2$} orbits,
perpendicular to it \citep{1980A&A....92...33C}. A double-barred potential needs an additional
frequency to be described, meaning that closed orbits are not possible. 
A loop is a family of orbits whose populating particles return
to the same curve when the two bars recover the relative position between them, 
although each individual particle is not in its initial position. This means that only the
global shape of the orbit is preserved.

The challenge of building 
long-lived double bars led to many
numerical simulations, most of them getting galaxies where the outer bar appears
first, the gas flows along it and then it is captured by the \emph{x$_2$} orbits of the main bar, 
subsequently forming the inner bar. In this scenario the dissipative component plays a major role
\citep{1993A&A...277...27F,2001ApJ...553..661H,2002ApJ...565..921S,2002MNRAS.337.1233R,2004ApJ...617L.115E}. 
However, some authors get double-barred systems without such a
dissipational process. This is the case of \citet{2007ApJ...654L.127D}, who perform collisionless N-body simulations 
and obtain purely stellar inner bars.
These inner bars form before the apparition of the outer bars.
Finally, \citet{2007ApJ...657L..65H} generate for the first time a two bar system 
from a dark matter halo. In this
case, both bars are gas-rich and transport material to the central
regions, supporting the hypothesis of \citet{1989Natur.338...45S,1990Natur.345..679S}.

With these pieces of evidence, it is clear that double-barred galaxies are 
the perfect benchmark to constrain the role of secular processes within galaxy nuclei.
One way to proceed is to perform a detailed stellar population analysis of
these systems: if double bars are playing a major role in galaxy evolution, this fact will
be reflected in the properties (star formation, age, and metallicity) of the galaxy components
(bulge, inner bar, outer bar, and disc),
resulting in a gradient of ages and metallicities between them.
However, this is difficult to achieve due to the structural complexity
of galaxies with nested bars, as illustrated by the photometric properties mentioned above. 
To date, there are few stellar population studies involving double bars, all of them being projects
devoted to the analysis of samples of barred galaxies in general, which include only some
double-barred objects. For this reason, these works focus only on the main bar and
they do not account for the inner bar. This is the case in the set of papers by 
\citet{2007A&A...465L...9P,2009A&A...495..775P}, \citet{2011A&A...529A..64P}, and \citet{2011arXiv1103.3796S},
who study the stellar populations of bars and
bulges. They find positive, negative, and even null metallicity gradients along the bars
independently of the age distribution. The bulges hosted
in  barred
galaxies tend to be more metal-rich than those in unbarred galaxies at a similar velocity
dispersion. However, the implications of these results within evolutionary scenarios
is still unclear.

Other works study double-barred objects 
from a purely kinematical point of view
\citep[][]{2001A&A...368...52E,2004A&A...421..433M,2008ApJ...684L..83D}. 
\citet{2008ApJ...684L..83D}
present the 2D stellar velocity and velocity dispersion maps for a sample of four double-barred
galaxies, based on observations with the SAURON integral field spectrograph. The analysis of
high quality velocity
dispersion maps revealed two local minima,
located exactly at the ends of the inner bar of each galaxy.
By means of numerical simulations, \citet{2008ApJ...684L..83D} conclude that these
\emph{$\sigma$-hollows} appear because of the strong 
contrast between the velocity dispersions of the
bulge and the inner bar. This result indicates that bars are dynamically cold components, with
velocity dispersion values significantly lower than those of the hot, classical bulge.
Moreover, the $\sigma$-hollows are the kinematical signature of the presence of an inner bar.

In this paper, we present the first detailed morphological, kinematical and stellar population
analysis of the different structures that comprise a double-barred galaxy: bulge, 
inner bar and outer bar. We focus on the case of NGC\,357. The work is organised as follows: in Section
\ref{sec:thegalaxy} we list the relevant properties of NGC\,357 from the literature, 
while in Section \ref{sec:obs} we summarise the observations and the data reduction 
procedure. Then, we describe the morphology
and kinematical profiles in Sections \ref{sec:mor} and \ref{sec:kin}, respectively.
Section \ref{sec:sp} details the stellar population analysis, from 
the emission lines correction of the spectra to the detailed study of the bulge region 
and the comparison between structural components.
A discussion of the implications
of these results for the structure and possible formation scenarios of NGC\,357 can
be found in Section \ref{sec:dis}. Finally, Section \ref{sec:con} sets out
the main conclusions of this work.

\section{Main properties of NGC\,357}\label{sec:thegalaxy}

\begin{figure*}
\begin{center}
  \includegraphics[angle=0,width=.8\textwidth]{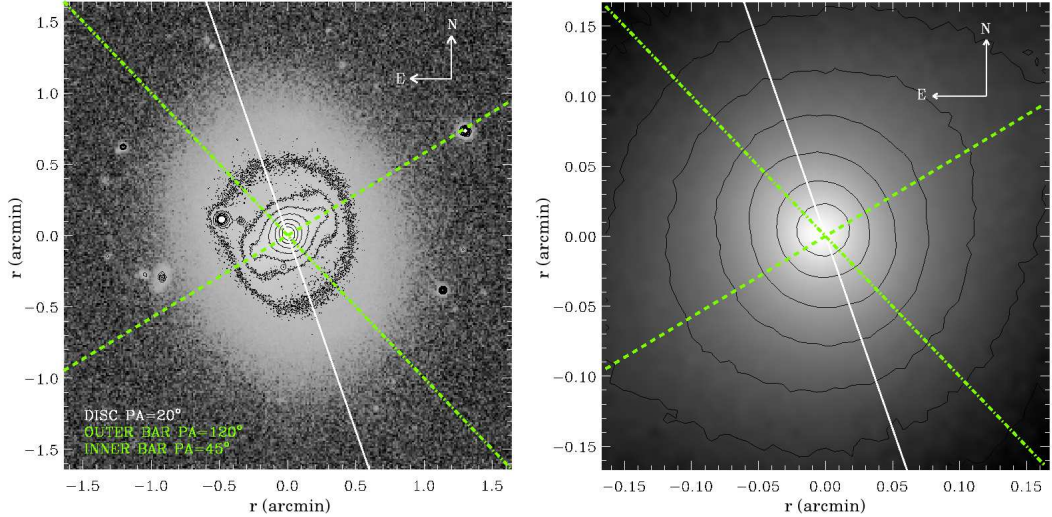}
  \caption{\emph{r}-band image of NGC\,357 from the Sloan Digital Sky Survey \citep{2000AJ....120.1579Y}.
    The left panel shows the whole galaxy whereas the right panel is a zoom of the inner $\sim$20 arcsec.
  We have overplotted the isodensity contours and the directions of the major axis of the inner bar
  (green dot-dashed line), outer bar (green dashed line) and disc (white line).}
  \label{fig:ima}
\end{center}
\end{figure*}

NGC\,357 is classified as SB(r)0/a by \citet{1991trcb.book.....D} and as SBa by \citet{1981RSA...C...0000S}.
This galaxy appears in the catalogue of double-barred galaxies of \citet{2004A&A...415..941E},
who provides information about the properties of the main and the inner bar of NGC\,357,
based on the analysis of infrared images from the Hubble Space Telescope.
The two bars of NGC\,357 are almost perpendicular, with 
the inner one almost parallel to the major axis of the galaxy, as seen in Figure \ref{fig:ima}.
We have specifically chosen this early-type 
spiral in order to avoid the presence of complex structural components and dust, which can make
the stellar population analysis very tricky. 
The galaxy distance is 31.6 Mpc \citep{2004A&A...415..941E}, which corresponds to $\sim$150 pc\,arcsec$^{-1}$.
The relevant properties of NGC\,357 are shown in Table \ref{tab:n357}. 

\begin{table}
\begin{center}
  \caption{Relevant properties of the structural components of NGC\,357.}
  \label{tab:n357}
  \begin{tabular}{ccccc}
    \hline
    \hline
\multicolumn{5}{c}{NGC\,357}\\
    \hline
          & Inclination  & PA           & Semi-major axis & $\epsilon_{max}$ \\
          & (1)          & (2)          & (3)             & (4)             \\\hline
Inner bar &      -       & 45$^{\circ}$  & 3.1 arcsec      & 0.16            \\
Outer bar &      -       & 120$^{\circ}$ & 21 arcsec       & 0.44            \\
Disc      &  36$^{\circ}$ & 20$^{\circ}$  & -               & -               \\\hline
    \end{tabular}
\end{center}
(1) galaxy inclination \citep{2005A&A...434..109A}; \\ 
(2) position angle (hereafter PA) of the major axis of the galaxy \citep[disc;][]{2005A&A...434..109A} 
and the two bars \citep{2004A&A...415..941E};\\ 
(3) bar lengths, estimated as the semi-major axes of maximum isophotal ellipticity \citep{2004A&A...415..941E}, and \\ 
(4) maximum isophotal ellipticity of the bars ($\epsilon=1-b/a$), from \citet{2004A&A...415..941E}. 
\end{table}

Although some studies claim this galaxy is isolated \citep[e.g.,][]{2007MNRAS.381..943G}, 
\citet{2002AJ....124..782V} finds it belongs to a group
with at least six other members. However, it shows no
signatures of interaction with its closer companions. NGC\,357 is classified as a LINER
\citep{2007MNRAS.381..943G}.

The structural properties of NGC\,357 have been studied photometrically by several authors.
In this way, \citet{2005A&A...434..109A} perform a photometric decomposition of
an \emph{I}-band image of this galaxy along its outer
bar major axis, getting the main photometric parameters of its structural components, namely
bulge, outer bar and disc; they do not take into account the inner bar since its contribution
to the radial surface brightness along the main bar is negligible.
Their main conclusion is that the bulge of NGC\,357 follows the same
fundamental plane than the ellipticals and other bulges of S0 galaxies.

\section{Observations and data reduction}\label{sec:obs}

The spectroscopic observations of NGC\,357 were carried out with the
3.5-m New Technology Telescope (NTT) at the European Southern
Observatory (ESO) in La Silla (Chile) on 5-13 October 2002.  The ESO
Multi-Mode Instrument (EMMI) was operated both in blue (BLMD) and red
medium-dispersion spectroscopic (REMD) mode.

The NTT mounted EMMI in BLMD using the grating No.\,3 blazed at 3800\,\AA\ 
with 1200 grooves\,mm$^{-1}$ in first order in combination with a
1.3 arcsec $\times$ 5.5 arcmin slit.  The detector was the No.\,31
Tektronix TK1024 EB CCD with $1024\times1024$ pixels of $24\times24$
$\mu$m$^2$. It yielded a wavelength coverage between about 3990\,\AA\ 
and 4440\,\AA\ with a reciprocal dispersion of 0.45\,\AA\ pixel$^{-1}$. 
The spatial scale was 0.37 arcsec\,pixel$^{-1}$.
The instrumental resolution was 1.3\,\AA\ (FWHM) corresponding to
$\sigma_{\rm inst}\sim40$ km s$^{-1}$.

Six spectra of 45 min each were taken aligning the slit with the
major axis of the inner bar ($\rm PA = 45^\circ$). Two more spectra
of 45 min each were obtained aligning the slit with the major axis of
the outer bar ($\rm PA = 120^\circ$).  All the spectra were obtained
using the guiding TV camera to center the slit on the galaxy nucleus.

The NTT mounted EMMI in REMD using the grating No.\,6 blazed at 5200\,\AA\ 
with 1200 grooves mm$^{-1}$ in first order. A 1.0 arcsec $\times$
5.5 arcmin slit was adopted. The mosaiced MIT/LL CCDs No.\,62 and 63
with $2048\times4096$ pixels of $15\times15$ $\mu$m$^2$ covered the
wavelength range between about 4800\,\AA\ and 5460\,\AA . The on-chip
$2\times2$ pixel binning provided a reciprocal dispersion and spatial
scale of 0.40\,\AA\ pixel$^{-1}$ and 0.332 arcsec pixel$^{-1}$. The
instrumental resolution was 1.6\,\AA\ (FWHM) corresponding to
$\sigma_{\rm inst}\sim30$ km s$^{-1}$.

After centering the slit on the galaxy nucleus, two spectra of 30
min each were obtained along the disc major axis ($\rm PA =
20^\circ$).

The range of the seeing FWHM during the observing runs was 0.6-1.4
arcsec as measured by the ESO Differential Image Motion Monitor.  A
comparison lamp exposure was obtained after each object integration to
allow accurate wavelength calibration. Quartz lamp and twilight sky
flatfields were used to remove pixel-to-pixel variations and
large-scale illumination patterns. Several G and K stars and
spectrophotometric standard stars were observed with the same set-up
to serve as templates in measuring the stellar kinematics and in flux
calibration, respectively.

All the spectra were overscan and bias subtracted, flatfield
corrected, corrected for bad pixels and columns, and wavelength
calibrated using standard IRAF\footnote{Imaging Reduction and Analysis
  Facilities (IRAF) is distributed by the National Optical Astronomy
  Observatories which are operated by the Association of Universities
  for Research in Astronomy (AURA) under cooperative agreement with
  the National Science Foundation.} routines. The cosmic ray removal
is a critical step since any residual might affect the spectral lines
measured to derive the properties of the stellar populations. It was
performed with the REDUCEME package \citep{1999PhDT........12C}, that
assures a careful and accurate inspection and interpolation of the
spectra.  We checked that the wavelength rebinning was done properly
by computing the difference between the measured and predicted
wavelengths \citep{1996PASP..108..277O} for the brightest night-sky
emission lines in the observed spectral ranges. The resulting accuracy
in the wavelength calibration is better than 5 km s$^{-1}$. The
spectra taken along the same axis were co-added using the centre of
the stellar continuum as reference. The contribution of the sky was
determined from the outermost regions at the two edges of the
resulting spectra, where the galaxy light was negligible, and then
subtracted, giving a sky subtraction better than 1\%. A
one-dimensional sky-subtracted spectrum was obtained for each
kinematical template star. The sky-subtracted spectra were flux
calibrated using the observed spectrophotometric stars as a reference.
Neither extinction nor dust corrections were applied for the blue
spectra since the wavelength coverage is short and we are not
interested in colour measurements.

Finally, the galaxy spectra were binned along the spatial direction in
order to assure a minimum signal-to-noise ratio (S/N hereafter) of $\sim20$
\AA$^{-1}$, sufficient to our kinematic analysis.
Such a minimum S/N is reached only in the outermost radial bins, it
increases at smaller radii reaching a maximum S/N$\sim60$ \AA$^{-1}$
in the radial bin corresponding to the galaxy centre. 
The spatial binning procedure was carefully carried out 
in order to assure
the separation of the different
structural components of NGC\,357 (i.e., the bulge, inner, and outer
bar).

\section{Analysis and results}\label{sec:aar}
\subsection{Photometry}\label{sec:mor}

\begin{figure*}
\begin{center}
  \includegraphics[angle=0,width=1.\textwidth]{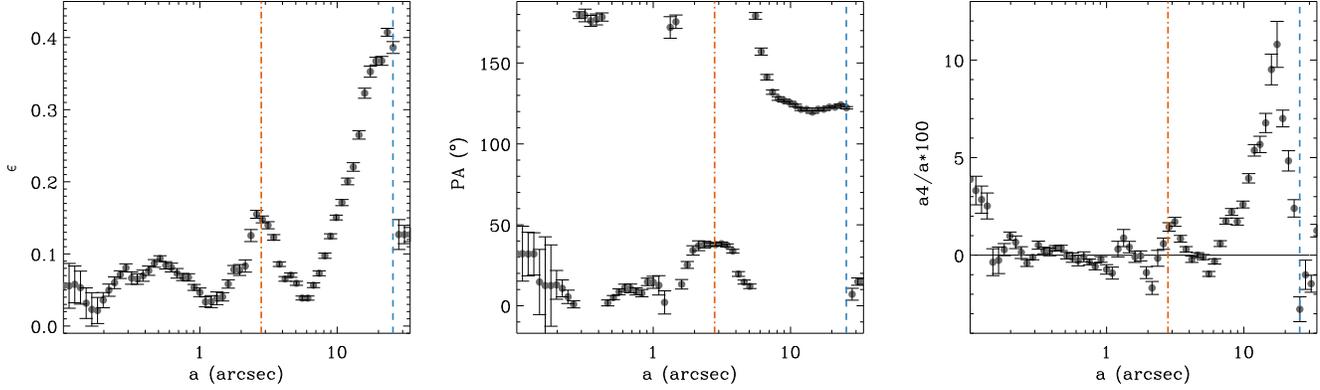}
  \caption{Ellipticity (left panel), position angle (middle panel), and fourth
    cosine Fourier coefficient (right panel) 
    profiles that result from fitting
    ellipses to the isophotes of NGC\,357 in the NICMOS2 F160W image. 
    The semi-major axes of the fitted ellipses are given in
    logarithmic scale.
    The vertical orange dot-dashed and blue dashed lines indicate the lengths of the inner 
    and outer bars, respectively.}
  \label{fig:mor}
\end{center}
\end{figure*}

As a first approach to disentangling the morphology of NGC\,357, we perform an ellipse
fitting over a NICMOS2 F160W image taken from the Hubble Legacy Archive 
(Prog. Id. 7330, PI John Mulchaey). For this purpose
we use the IRAF task ELLIPSE, which follows the procedure outlined by \citet{1987MNRAS.226..747J}.
Figure \ref{fig:mor}
shows the resulting profiles of ellipticity, $\epsilon$, position angle, PA, 
and the fourth cosine Fourier coefficient, $\emph{a}_4$.
The presence of the bars causes two sharp changes in the ellipticity and PA parameters: one peak at 
$\sim$3 arcsec related to the inner bar, and a discontinuity at $\sim$25 arcsec
due to the outer bar. These two values are in agreement with the bar lengths
estimated by \citet{2004A&A...415..941E}. The $\emph{a}_4$ profile also shows two peaks
at the locations of the two bars. $\emph{a}_4$ is a measurement of the deviation
of the isophotes from pure ellipses, so positive values correspond to discy isophotes
whereas negative values to boxy isophotes \citep{1987MNRAS.226..747J,2009ApJS..182..216K}. 
Therefore, the two bars of NGC\,357 present
discy isophotes.

We get a maximum isophotal ellipticity of $\epsilon\sim$0.15 and $\epsilon\geq$0.4
for the inner and outer bars, respectively, and a
PA$\sim$40$^{\circ}$ for the inner bar,
PA$\sim$120$^{\circ}$ for the outer bar, and PA$\sim$15$^{\circ}$ for the disc.
These measurements are also consistent with the estimates of 
\citet{2004A&A...415..941E} and \citet{2005A&A...434..109A}.

\subsection{Kinematics}\label{sec:kin}

\subsubsection{Measuring the line-of-sight velocity and velocity dispersion}

Two different techniques were used to derive the velocity and velocity dispersion profiles of
NGC\,357: a cross-correlation in the Fourier domain and a 
full spectrum fitting in wavelength space.
The cross-correlation procedure \citep[e.g.,][]{1979AJ.....84.1511T} is as follows: 
among the observed velocity standards, we
select those that are spectroscopically most similar to the galaxy. After
correcting them to rest frame, we combine the stellar spectra to create an artificial velocity standard
star spectrum that will be used as a template for the cross-correlation.
This standard spectrum is then broadened to several velocity dispersions and the results are
correlated with the original star, obtaining for each case a cross-correlation peak with its
full width half maximum (FWHM). Thus, we create a calibration table that provides the FWHM
we expect if a spectrum with a given velocity dispersion is cross-correlated with our stellar template, so
we can get the dispersion of the galaxy spectra by correlating them with the star.
The line-of-sight velocity is a direct product of the correlation since the template is
at rest.

For the full spectrum fitting we have made use of the penalized pixel fitting 
(hereafter pPXF) routine developed by \citet{2004PASP..116..138C}. The spectral regions which 
are potentially affected by emission lines are previously masked on each galaxy absorption spectrum, 
which is then fitted with
a linear combination of a well-selected subsample of the single
stellar population models (hereafter SSPs) 
of \citet{2010MNRAS.404.1639V}, previously convolved
with a line-of-sight velocity distribution (hereafter LOSVD). This LOSVD is parametrized by 
an expansion in Gauss-Hermite
functions \citep{1993MNRAS.265..213G,1993ApJ...407..525V}; since the S/N requirements for this procedure 
get more demanding as the order of the desired LOSVD moments increases, we decided to focus only
on the velocity and velocity dispersion to avoid the loss of spatial resolution 
caused by a wider radial binning.

The two methods used for deriving the kinematics of NGC\,357 are completely independent and even
work in different parameter spaces. In order to enhance the differences between both techniques, we 
explore two different ways that offer
complementary advantages: whereas the SSPs help to alleviate the template
mismatch problem \citep{2003A&A...405..455F},
the artificial template was obtained by stellar spectra, which were 
observed and reduced exactly in the same way as the galaxy spectra, thus
minimizing the possibility of introducing instrumental effects.

The results obtained from both methods
are fully consistent, with differences smaller than the corresponding error bars. These differences range
from few km\,s$^{-1}$ in the central regions to a maximum of 10 km\,s$^{-1}$ in the farthest observed radii.
Figure \ref{fig:kin} shows the final line-of-sight velocity (after subtracting the systemic velocity)
and velocity dispersion
profiles obtained with pPXF along the major axis of the inner bar, outer bar, and disc.

\subsubsection{Revealing the inner and outer bars}

\begin{figure*}
\begin{center}
  \includegraphics[width=\textwidth]{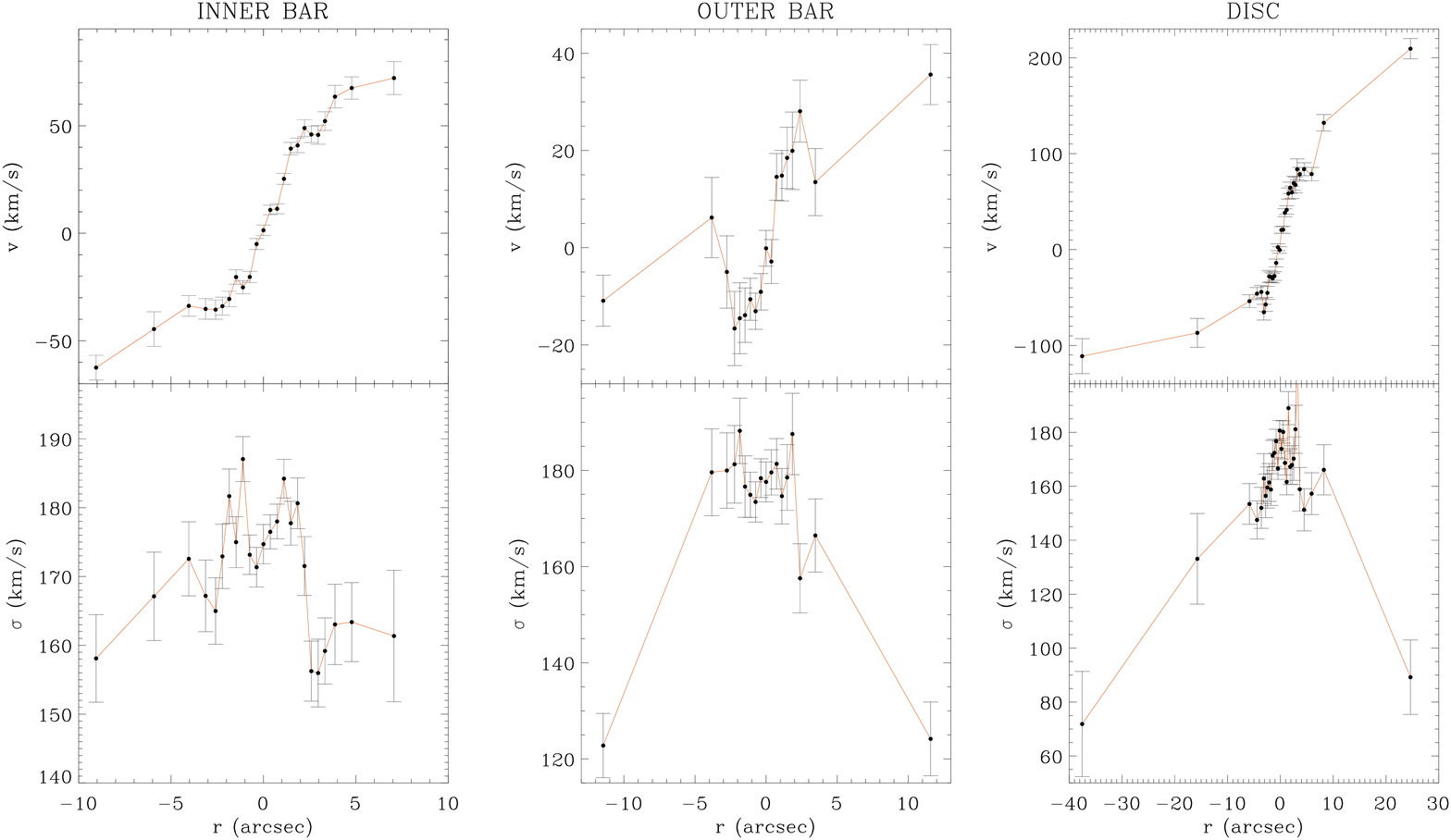}
  \caption{Line-of-sight velocity (after the systemic velocity subtraction; top panels) 
    and velocity dispersion (bottom panels) profiles as a function of the radius for the spectra
    obtained along the major axis of the inner bar (left panels), outer bar (middle panels) and disc
    (right panels) of NGC\,357.}
  \label{fig:kin}
\end{center}
\end{figure*}

The presence of the inner bar is clearly unveiled by the shape of the rotation curve
(Figure \ref{fig:kin}, left panels), that follows 
the same characteristic \emph{S-} or \emph{double-hump} profile previously observed for
inner bars by \citet{2001A&A...368...52E}. 
This profile is characterized by an inner
steep slope so the rotation reaches a local maximum, it slightly drops to a local minimum
and then it raises again slowly further out. Since the two bars of NGC\,357 are almost perpendicular,
with the inner one almost parallel to the major axis of the galaxy,
we can conclude that the double-hump profile in Figure \ref{fig:kin} is completely due to the 
presence of the inner bar. This kind of rotation curve was theoretically derived by 
\citet{2004AJ....127.3192C} and
\citet{2005ApJ...626..159B} by means of N-body simulations of single-barred galaxies; in this last work,
the authors show that the double-hump structure is more enhanced in galaxies
with strong bars seen end-on, whereas the effect is weakened for
larger viewing angles or weaker bars. Although the double-hump rotation curve
is clearly visible in the case of the inner bar of NGC\,357, suggesting the bar is strong,
the maximum isophotal ellipticity is only $\epsilon_{max}\sim$0.15, 
which indicates that it is actually a very weak bar. Indeed,
the maximum ellipticity of the isophotes
in a bar region correlates well with the bar strength \citep{2002MNRAS.331..880L}.

The velocity dispersion profile along the inner bar shows three local minima:
a central drop corresponding to the so-called \emph{$\sigma$-drop}
of \citet{2001A&A...368...52E}, and two symmetric decreases exactly at the edges of the bar ($\pm$4 arcsec) 
that can be identified
as the \emph{$\sigma$-hollows} \citep{2008ApJ...684L..83D}, known to be signatures 
of an inner bar. These hollows have an amplitude of $\sim$20 km\,s$^{-1}$,
consistent with the findings of \citet{2008ApJ...684L..83D}, who found amplitudes ranging from
10 to 40 km\,s$^{-1}$.
Further discussion on these
features and their different origins 
is found in Section \ref{sec:dis}. As expected, the velocity dispersion decreases outwards.

The outer bar of NGC\,357 (Figure \ref{fig:kin}, middle panels)
is almost perpendicular to the major
axis of the galaxy, which we assume to be the kinematical line of nodes. This means
that the line-of-sight velocity values along the outer bar
are expected to be very low, as it is shown in Figure \ref{fig:kin}.
However, a clear kinematical decoupled 
profile appears in the central region. This structure rotates faster than its surroundings and
extends $\sim\pm$2 arcsec, which corresponds to the radial region of the $\sigma$-drop observed in the
velocity dispersion profile. 
A closer inspection to the inner bar velocity profile reveals the same decoupling, since two steep humps
appear inside the double-hump profile, exactly
at $\sim\pm$2 arcsec. 
Unfortunately, the low S/N level of the spectra along 
the direction of the outer bar of NGC\,357 only allows us to
analyse the kinematics in the inner $\sim\pm$10 arcsec, so we cannot explore the whole main bar
to see if other signatures, as for example the $\sigma$-hollows or the double-hump profile, are present. 

The kinematics along the major axis of the disc is shown in Figure \ref{fig:kin} (right panels).
In this case, we find as expected the maximum 
line-of-sight velocity values and the same central signatures found for the other directions:
the decoupling
in the velocity profile and the $\sigma$-drop. In Section \ref{sec:dis} we discuss the possible 
structures that might be causing these signatures.

\subsection{Stellar populations}\label{sec:sp}

The aim of this stellar population analysis is to recover the relative ages,
metallicities, and formation timescales of the different structural components of
NGC\,357: bulge, inner bar, and outer bar. Unfortunately, the quality of the data,
although very high, still is not sufficient to analyse the disc, which requires
spectra taken with 10 m-class telescopes. To perform this detail study, we have
summed up the individual spectra, previously corrected for their kinematics 
(velocity and velocity dispersion), to reach the desired S/N within each structural
component. For this purpose, we only add
those spectra corresponding to the regions
where a given component is clearly dominating the total luminosity: the inner $\pm$1 arcsec for the bulge, 
from $\pm$2 arcsec to $\pm$4 arcsec for the inner bar, 
and from $\pm$4 arcsec to $\sim\pm$15 arcsec
(limited in this case by S/N requirements) for the outer bar.
Since the spectra along the major axes of the two bars and the disc always cover the central region,
we have two spectra for the bulge in the blue and red spectral range, respectively. However, the bars were
observed only in the blue spectral range. 

Again, we have chosen two methods to analyse the stellar populations:
the line-strength indices and a full spectrum-fitting approach.
Unlike using the Lick/IDS system, we have chosen to perform our line-strength
analysis on the system imposed by the kinematics of the galaxy, as done in
\citet{2001ApJ...551L.127V} and \citet{2006ApJ...637..200Y}.
For the second approach, we use the cross-correlation technique
employed by \citet{1999ApJ...513..224V} and \citet{1999ApJ...525..144V}.

In both cases, comparisons
with stellar population synthesis models are needed; as for the kinematics, we have used the empirical 
models of \citet{2010MNRAS.404.1639V}, built
from the MILES stellar library \citep{2006MNRAS.371..703S}. These models have a wide
wavelength coverage (from 3525 to 7500 \AA) and a resolution of 2.5\,\AA\,(FWHM), 
constant over the whole spectral range \citep{2011A&A...531A.109B,2011A&A...532A..95F}.

\subsubsection{Emission line correction}\label{sec:emission}

The ionized gas present in the galaxy
leads to emission lines that contaminate the spectrum and fill in the
absorption lines. Therefore, it is crucial to correct for this emission to get the properties
of the stellar populations from the spectra. Indeed, 
if a proper separation of both contributions (the 
stellar absorptions and gaseous emissions) is not performed before the analysis,
the measured ages will be artificially older. In our particular case, we need to clean up
the three Balmer lines included in our spectral range (H$\beta$, H$\gamma$ and H$\delta$),
and the Fe5015 and Mgb indices, which can be affected by
the [OIII]4959,5007 and [NII]5198,5200 lines, respectively.

The common procedure for this correction is to mask the spectral region possibly
contaminated by emission lines and fit the obtained spectrum with a set of stellar population
models. For this purpose we use the ULySS\footnote{Available at http://ulyss.univ-lyon1.fr/.} package
\citep{2009A&A...501.1269K}, which minimizes the $\chi^2$ between the observed
spectrum and a model spectrum consisting of a linear combination of non-linear parameters (age and
metallicity), corrected for the kinematics and multiplied by a polynomial to avoid flux calibration
problems. We subtract the ULySS best fit spectrum to the original one in order
to get the residuals plus the contribution of the emission lines. We then approximate each emission line with a Gaussian,
considering only those features with an amplitude larger than three times the standard deviation of the residuals.

Given the sensitivity
of the emission correction, we double-check it by repeating it with
the GANDALF package \citep{2006MNRAS.366.1151S}. The advantage
of GANDALF is that it fits simultaneously the stellar and the gaseous contributions, 
by including emission lines as Gaussians in the set of stellar population models 
used as templates. Although GANDALF is also based on $\chi^2$ minimization, it differs from 
ULySS in that it fits the stellar contribution 
with a linear combination of models, while ULySS interpolates the grid of templates to look for the most
appropriate single stellar population model.

\begin{figure*}
\begin{center}
  \includegraphics[width=\textwidth]{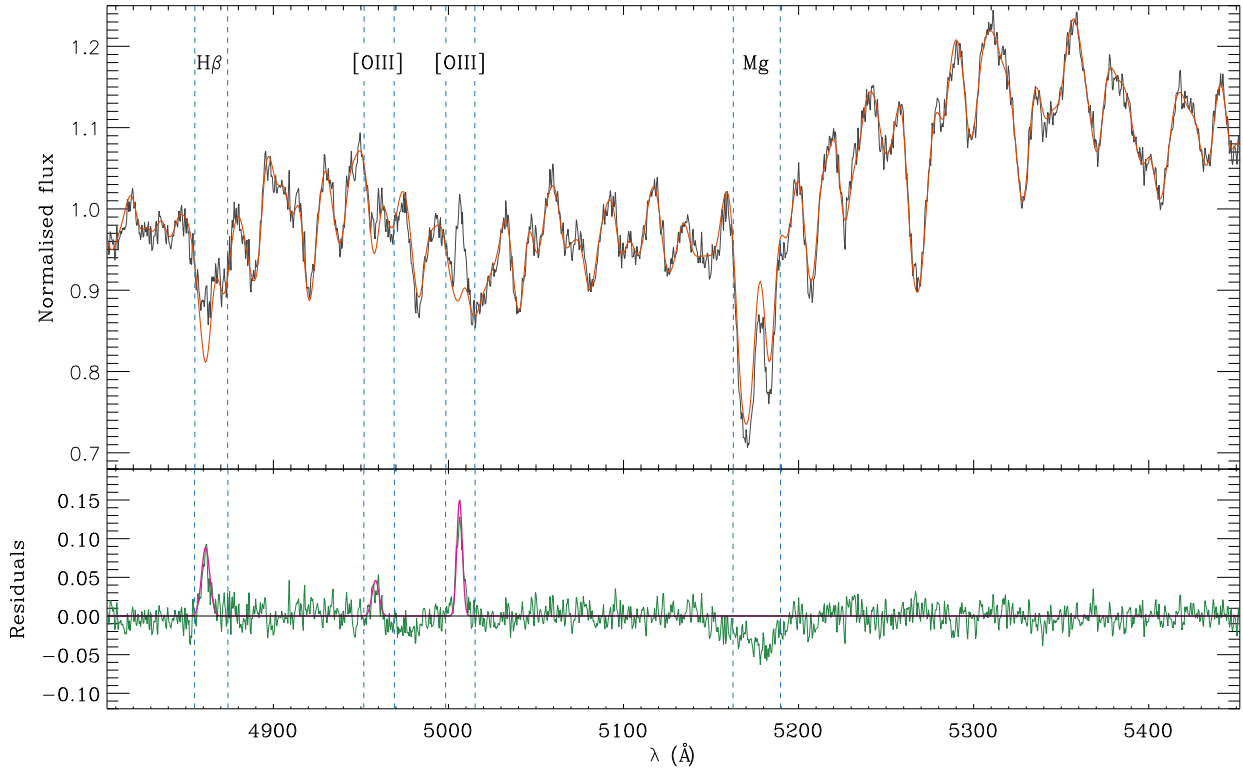}
  \caption{The correction of the emission lines in the red spectrum of the bulge of NGC\,357 by means of 
  ULySS. The observed and model spectra are plotted in black and orange, respectively in the upper panel. 
  The vertical dashed lines mark the spectral regions we masked because of 
  emission-line contamination (H$\beta$ and [OIII]4959,5007 doublet) and $\alpha$-enhancement (Mg).
  The residuals of the fit are shown in green  in the lower panel. The Gaussians fitting the emission 
  lines are plotted in magenta.}
  \label{fig:emission}
\end{center}
\end{figure*}

There are many details to take care of when fitting the full spectrum, such as
the initial guesses and the number and range of the masked regions. 
For example, in the case of the blue spectra of NGC\,357, we
are forced to mask the CN band as we find evidences of its $\alpha$-enhancement \citep[not included in the 
scaled-solar models of ][]{2010MNRAS.404.1639V}. 
When the amount of emission derived from the two routines is in agreement,
we subtract it from the observed spectra and use the results for the analysis
of the stellar populations.
Figure \ref{fig:emission} presents an example of the fitting performed with ULySS for the bulge red spectrum because it shows 
H$\beta$ and [OIII]5007, which are the most prominent emission lines we observed. 

\subsubsection{Line-strength indices}\label{sec:indices}

\begin{figure*}
\begin{center}
  \includegraphics[angle=90,width=.8\textwidth]{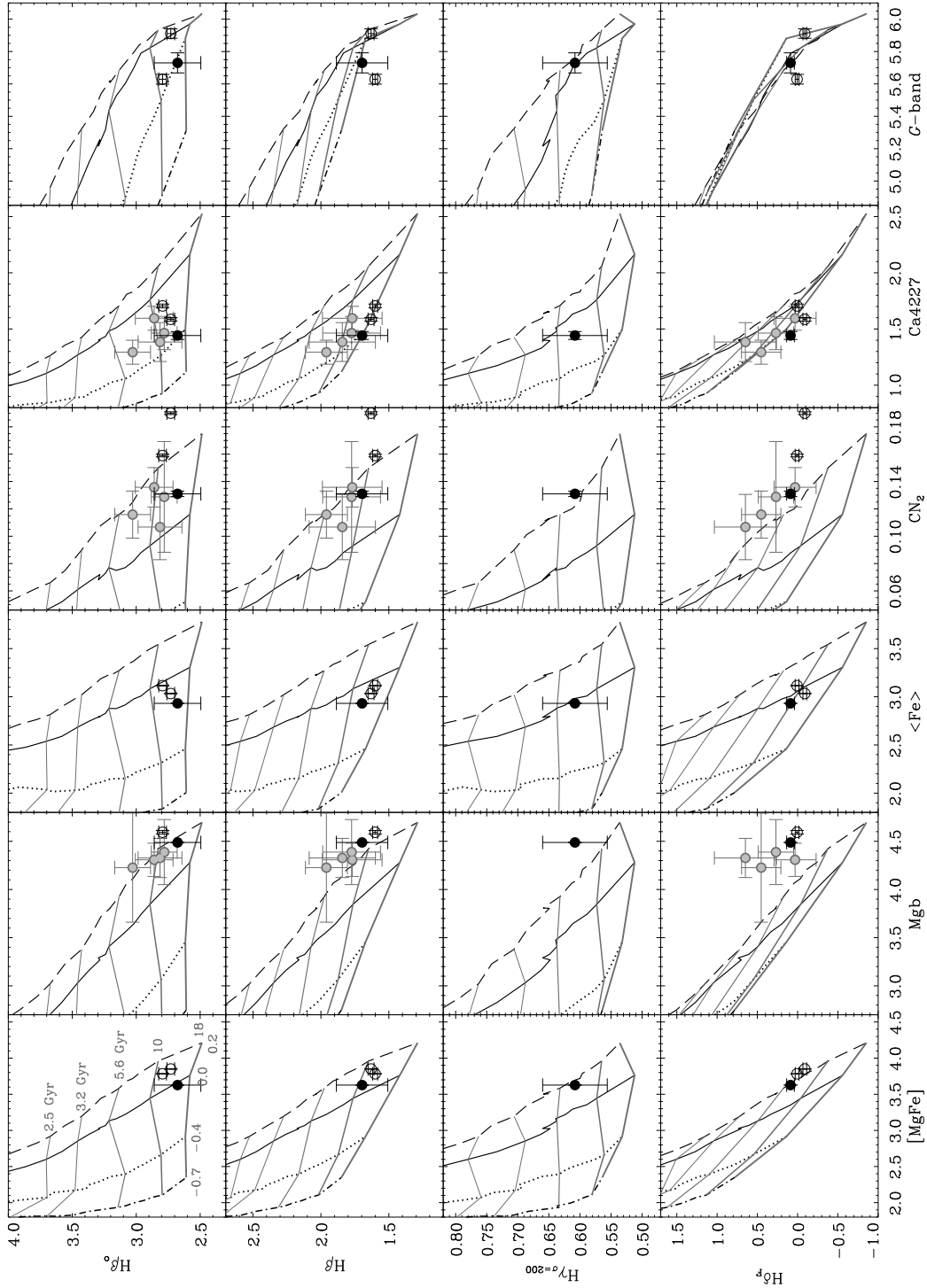}
  \caption{Age indicators H$\beta_o$, H$\beta$, H$\gamma_{\sigma=200}$, and H$\delta_F$
  (from top to bottom) versus the metallicity indicators [MgFe], Mgb, $\langle$Fe$\rangle$, 
  CN$_2$, and Ca\,4227, and the \emph{G}-band (from left to right).
  The grids correspond to the SSP models of \citet{2010MNRAS.404.1639V} once smoothed to match
  the resolution of the data ($\sigma\sim$180 km\,s$^{-1}$). The lines represent different
  ages increasing from top to bottom (2.5, 3.2, 5.6, 10, and 18 Gyr, respectively) 
  and metallicities, which  increase from left to right ($\rm [Z/H] = -0.7$, $-0.4$, 0.0, and 0.2, respectively).
  Ages and metallicities for the grids are indicated in the first panel.
  The black filled circle is the measurement for the bulge of NGC\,357, whereas the grey filled 
  circles are the mean values for the bulges of a subsample of barred galaxies from
  \citet{2009A&A...495..775P} and \citet{2011A&A...529A..64P}.
  The open circles correspond to two elliptical galaxies in Virgo cluster from \citet{2006ApJ...637..200Y}.
  All the reference galaxies have been selected to have a central velocity dispersion 
  similar to that of NGC\,357 ($\sigma\sim$180 km\,s$^{-1}$).}
  \label{fig:spBB}
\end{center}
\end{figure*}

\begin{figure*}
\begin{center}
  \includegraphics[width=.9\textwidth]{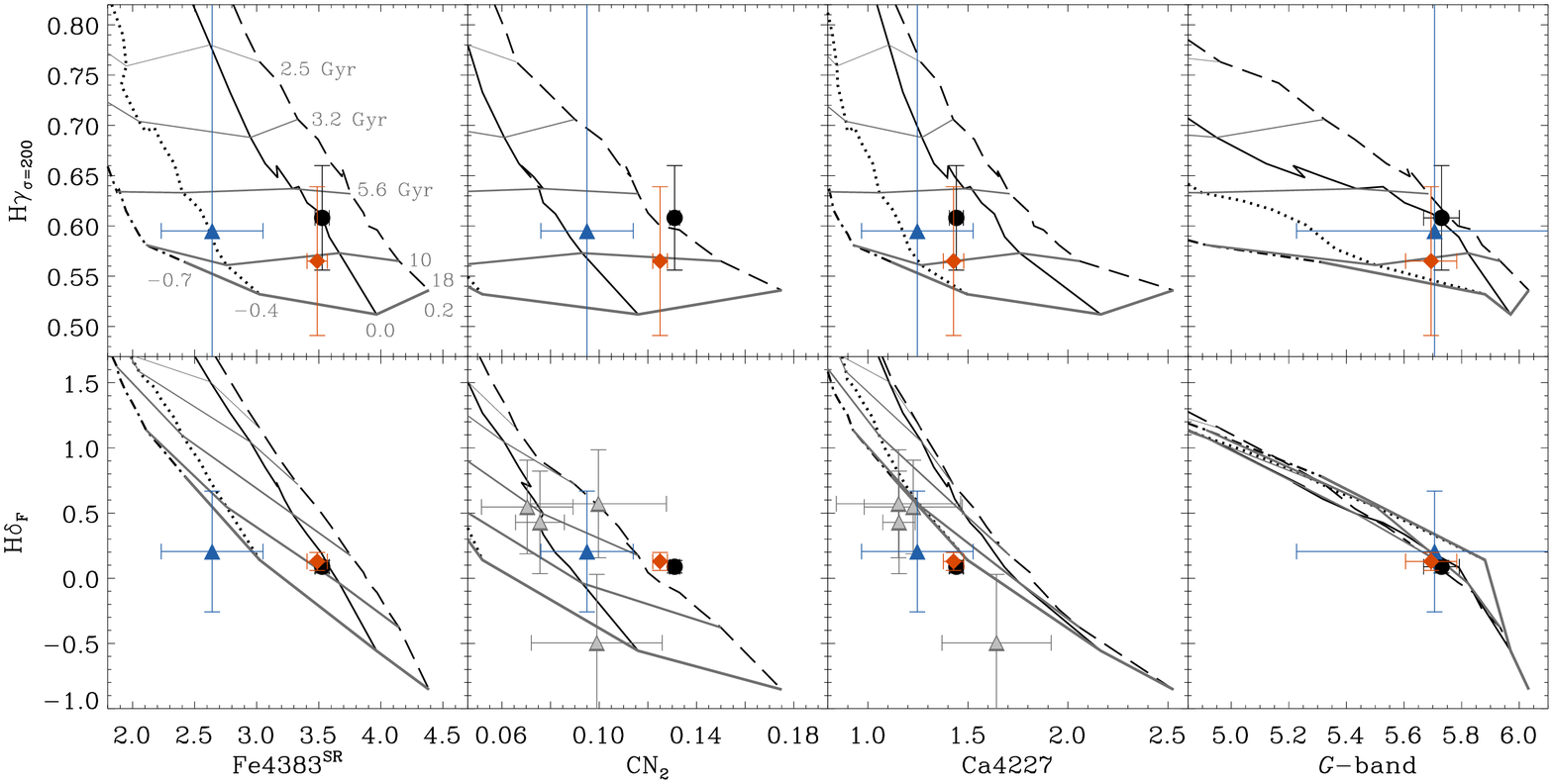}
  \caption{Age indicators H$\gamma_{\sigma=200}$ and H$\delta_F$
  (top and bottom panels, respectively) versus the metallicity indicators Fe4383$^{SR}$, CN$_2$, 
  and Ca\,4227, and the \emph{G}-band
  (from left to right).
  The model grids are the same as in Figure \ref{fig:spBB}.
  The coloured symbols are the measurements for the different structural components of NGC\,357:
  bulge (black circle), inner bar (orange diamond), and outer bar (blue triangle).
  The grey triangles are the mean values for the main bars of a subsample
  of galaxies from \citet{2009A&A...495..775P}, selected to have a 
  similar central velocity dispersion than NGC\,357 ($\sigma\sim$180 km\,s$^{-1}$).}
  \label{fig:spALL}
\end{center}
\end{figure*}

A set of suitable absorption line-strength indices are measured over the emission-cleaned spectra
of the bulge, inner bar, and outer bar, as well as 
over the \citet{2010MNRAS.404.1639V} models. In order to compare the results
of the three components, the galaxy spectra and the SSPs are 
first degraded to their maximum resolution, given by 
the higher value of the velocity dispersion profile ($\sigma\sim$180 km\,s$^{-1}$). 
All the measurements are performed in the same system, i.e., the one defined by the kinematics of
the galaxy in a consistent way. We
plot the values obtained for an age indicator versus those of a metallicity indicator for the SSP models,
in order to obtain model grids such as those shown in Figures
\ref{fig:spBB} and \ref{fig:spALL}. These grids are not perfectly
orthogonal due to the age-metallicity degeneracy. The mean luminosity-weighted 
ages and metallicities of each structural component of NGC\,357
can then be derived by simply overplotting the observed indices on these grids.

In particular, we computed all the
Lick indices included in our spectral range, following the definitions given by 
\citet{1998ApJS..116....1T}. 
Note that although the index definitions are similar, our model measurements are not
in the Lick/IDS resolution and IDS instrumental system, but at constant
$\sigma$ and flux-calibration response curve.
Since the red pseudo-continuum of the Fe4383 index falls
partially out of our blue spectral range, we have defined 
a new Fe4383$^{SR}$ index (named after \emph{Short Red}, see Appendix \ref{sec:apFe} for details). 
Table \ref{tab:fe} presents the definitions of the new Fe4383$^{SR}$ and the original
Lick/IDS Fe4383 index.

\begin{table*}
  \caption{ The definitions of the new Fe4383$^{SR}$ and Lick/IDS Fe4383 indices.}
  \label{tab:fe}
  \begin{tabular}{ccccc}
    \hline
    \hline
    Index & Blue pseudo-continuum  & Main bandpass     & Red pseudo-continuum & Source                      \\
          & (\AA)                  & (\AA)             & (\AA)                &                             \\
    \hline
    Fe4383$^{SR}$ & 4359.125-4370.375 & 4369.125-4398.087 & 4419.328-4432.389    & this paper                  \\
    Fe4383       & 4359.125-4370.375 & 4369.125-4420.375 & 4442.875-4455.375    & \citet{1998ApJS..116....1T} \\
    \hline
    \end{tabular}
\end{table*}

For the bulge red spectrum, we compute the combined Iron-index $\langle{\rm Fe}\rangle$
and total metallicity indicator [MgFe]\footnote{
$\langle{\rm Fe}\rangle = ({\rm Fe5270}+{\rm Fe5335})/2,\,{\rm and}$
$[{\rm MgFe}] = \sqrt{{\rm Mgb}\,\times \langle{\rm Fe}\rangle}$}
\citep{1993PhDT.......172G}. 
We avoid the use of the Fe5015 index, since it is highly contaminated 
by the prominent [OIII]5007 emission line.

In addition to the Lick age index-definitions for the
Balmer lines (H$\beta$, H$\gamma$ and H$\delta$), we also measure the
more powerful H$\beta_o$ \citep{2009MNRAS.392..691C} and H$\gamma_\sigma$ 
\citep{1999ApJ...525..144V} indices, which show great abilities to disentangle
ages. The use of indices with enhanced sensitivity to the age or metallicity
represents an advantage to lift the degeneracy between these parameters.

\subsubsection{Cross-correlation analysis}\label{sec:cc}

The alternative technique we use to analyse the stellar populations of NGC\,357 is 
to cross-correlate
the galaxy spectrum with each SSP spectrum from the \citet{2010MNRAS.404.1639V} model library.
For this purpose the model spectra are
previously prepared to match the spectral range, velocity dispersion, and spectral resolution of the data.
Moreover, the galaxy and model spectra are rebinned logarithmically 
and normalized to remove the continua.

In order to optimize the cross-correlation method
for disentangling different stellar populations, it is necessary to adequately
filter the spectra and multiply them by a cosine-bell-like function \citep{1979AJ.....84.1511T}.
The importance of choosing a suitable filter lies in 
the possibility of getting rid of the noise in the spectrum; this purpose can be
achieved by simply removing the largest wavenumbers, where the information
about the shortest wavelength ranges is included.
Therefore, the limit is imposed by the resolution of the data.
On the other hand, shorter wavenumbers contain information about
wider spectral ranges, so possible residuals of the continuum removal
due to errors in the flux calibration might also be filtered. The drawback of this procedure is that 
it implies a power loss of the final cross-correlation function,
specially when filtering short wavenumbers where most of the signal is included.

Apart from the filtering, it might be required to mask some regions in the original
spectra, as is usually done in the full-spectrum fitting technique in the wavelength
space. We tested different masks trying to avoid those features that are 
not well reproduced by the models due to mismatched abundance ratios:
the CN in the blue spectral range and the Mg and
H$\beta$ features in the red spectral range. Again, these features contain most of the signal of the power spectrum
so the choice of the masks has to be done very carefully in order to not lose most of the
information.

Finally, the peak height obtained for each correlation function is plotted against the model age and 
metallicity. Since the cross-correlation profile reaches a higher value
when object and template are more similar (getting a maximum value of 1 if they are exactly the same
and no filtering is applied), 
the maximum value in the final peak heights plot indicates which template best resembles the
galaxy spectrum. This method takes advantage of the full information contained in the
spectrum, instead of being constrained to certain features as in the case
of the line-strength indices. This type of full-spectrum fitting technique
works in the Fourier space and, as an additional advantage, it does 
not need for a previous emission correction of the data. 
The application of this technique to the derivation 
of the stellar populations was already introduced by 
\citet{1999ApJ...513..224V} and \citet{1999ApJ...525..144V} and in this work it is adopted to
double-check the results obtained from the analysis of the line-strength indices.

\begin{figure*}
\begin{center}
  \includegraphics[width=\textwidth]{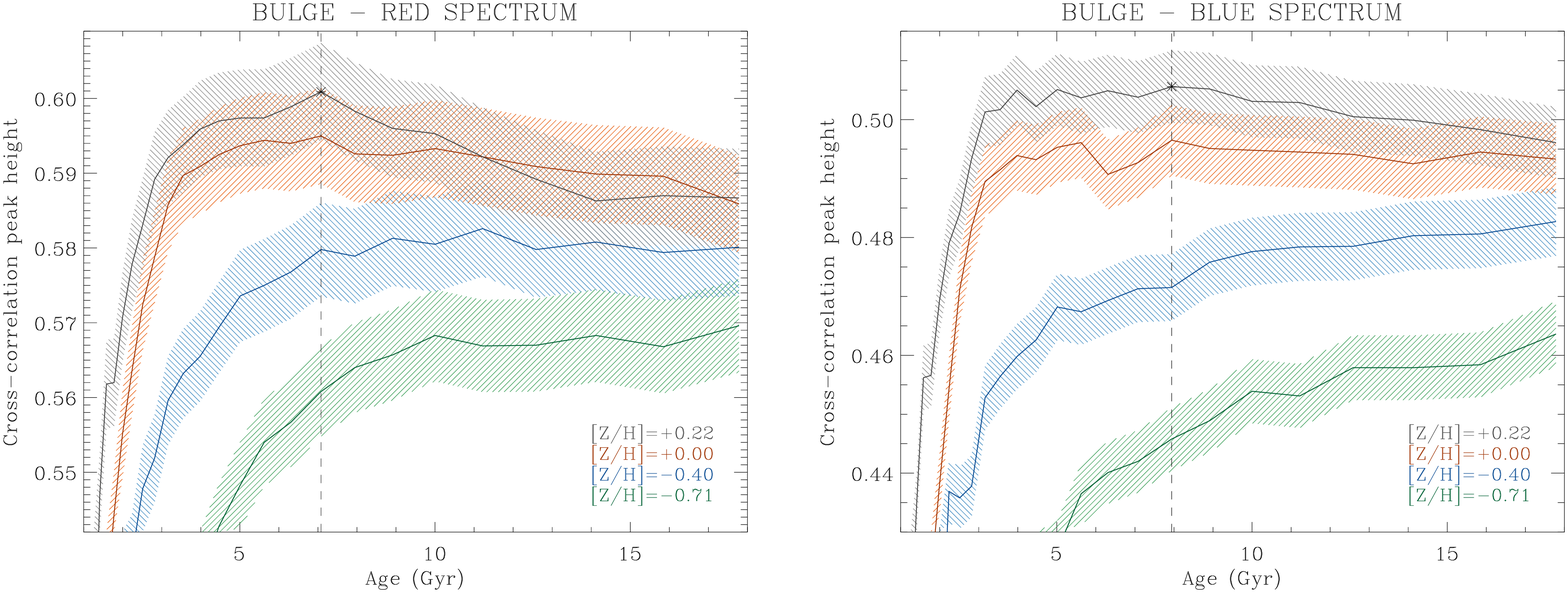}
  \caption{Cross-correlation peak heights from the analysis of the red (left panel) 
  and blue (right panel) spectra of the bulge of NGC\,357 
  with a set of SSP models of \citet{2010MNRAS.404.1639V}, 
  once smoothed to match the spectral range, 
  resolution, and dispersion of the data. The different coloured lines correspond 
  to metallicities of $\rm [Z/H]  = 0.2$ (black), 0.0 (red), $-0.4$ (blue), and $-0.7$ (green)
  as a function of age. The width of the coloured region represents the 1-$\sigma$ uncertainty 
  derived from Monte Carlo simulations. The largest cross-correlation peak height
  value in each plot is marked by an asterisk and the corresponding age with a vertical dashed line.}
  \label{fig:ccBB}
\end{center}
\end{figure*}

\begin{figure*}
\begin{center}
  \includegraphics[width=0.5\textwidth]{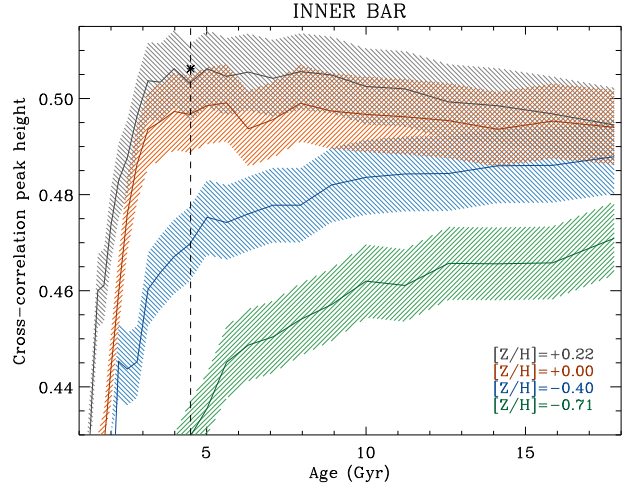}
  \caption{Same as in Figure\ref{fig:ccBB} for the blue spectrum of the inner bar of NGC\,357.}
  \label{fig:ccIB}
\end{center}
\end{figure*}

To compute the errors for the cross-correlation method 
we perform a set of 100 Monte Carlo simulations
by randomly adding Gaussian noise to the data, keeping the original S/N ratio.
The 1-$\sigma$ regions are then overplotted to the cross-correlation peak height curves. 
In order to test the requirements of this method we run some simulations with 
the \citet{2010MNRAS.404.1639V} models by artificially changing the different 
parameters involved: spectral resolution, spectral range, and S/N ratio,
taking as a reference the values measured for our observed spectra.
We find that the S/N of the data obtained for the outer bar is not enough to 
disentangle its age and metallicity, since the error bars for this case are 
too large (see Appendix \ref{sec:apcc}). This prevents us from providing a more constrained solution for
the outer bar through the cross-correlation analysis.
We have to note that the results obtained here are aimed to be complementary to those from the
line-strength analysis.
Therefore, we restrict the application of the cross-correlation to 
the bulge and inner bar. These results are shown in Figures \ref{fig:ccBB} and \ref{fig:ccIB}, and
will be analysed in Sections \ref{sec:bulge} and \ref{sec:gradients}, respectively.

\subsubsection{Stellar populations of the bulge}\label{sec:bulge}

Figure \ref{fig:spBB} shows the analysis of the line-strength indices of the bulge of
NGC\,357. Four age indicators 
are plotted against several well selected metallicity indicators
and the \emph{G}-band, using features from both the red and blue spectral ranges
available for this component. Balmer lines might be affected by emission line
contamination, although the amplitude of this emission decays from 
the top (H$\beta_o$ and H$\beta$) to the bottom (H$\delta_F$)
indices of Figure \ref{fig:spBB}. In fact, we find no emission in the H$\delta$ feature for the bulge.
For the case of H$\beta_o$ and H$\beta$, where the emission contamination
is larger, we
find some disagreements in the resulting ages with respect to those derived from H$\gamma$ and
H$\delta$. Although the emission correction has been performed very carefully (see Section
\ref{sec:emission}), these inconsistencies are probably due to an
underestimation of the emission affecting the H$\beta$ feature within the noise level. As an attempt to take
this effect into account, the H$\beta$ and H$\beta_o$ error bars
are obtained by computing the maximum and minimum (but no null) 1-$\sigma$
emission correction.

Moreover, we find subtle differences between the ages obtained with the H$\beta_o$ and H$\beta$ indices. These
differences might be attributed to either the emission correction, that affects
the two definitions in a different manner, or to the greater sensitivity of H$\beta_o$ to the
$\alpha$-enhancement in comparison to that of H$\beta$ \citep{2009MNRAS.392..691C}.
Indeed, these differences are more evident in the $\alpha$-enhanced indicators (Mgb and CN$_2$), whereas
the results from $\langle$Fe$\rangle$ or Ca\,4227 are almost identical.
Taking into account the error bars and the results from all the plots in Figure \ref{fig:spBB},
we derive a mean luminosity-weighted age of $\sim$8 Gyr for the bulge
of NGC\,357.

From the first column in Figure \ref{fig:spBB}, where the total metallicity indicator [MgFe]
is plotted, we derive a supersolar metallicity $\sim$0.2 dex for the bulge
of NGC\,357. For this component we also measure the $\langle$Fe$\rangle$,
Mgb, and CN$_2$ indices, so that it is possible to estimate the 
[Z$_{\rm Mg}$/Z$_{\rm \langle Fe \rangle}$] and [Z$_{\rm CN_2}$/Z$_{\rm \langle Fe \rangle}$] ratios,
which are calculated as the relative metallicity difference obtained when plotting the various metallicity
indicators versus the Mg \citep{2001ApJ...551L.127V}.
[Z$_{\rm Mg}$/Z$_{\rm \langle Fe \rangle}$] and [Z$_{\rm CN_2}$/Z$_{\rm \langle Fe \rangle}$]
are shown to be good proxies for the [Mg/Fe] and [CN/Fe] abundance ratios.
In fact, [Z$_{\rm Mgb}$/Z$_{\rm \langle Fe \rangle}$] and [Mg/Fe] follow a linear relation,
as shown in \citet{2010MNRAS.404.1639V}.
These overabundances are important because they
can be used to constrain the timescale for the formation of the bulk
of their stellar populations:
Mg (as well as other $\alpha$
elements) is the product of type II supernovae, which explode soon after the 
star formation; on the contrary, the Iron-peak elements are ejected by type Ia supernovae,
on longer timescales of around 1 Gyr. In the meantime between those two cases, the low- and 
intermediate-mass stars enrich the interstellar medium with C and N. Therefore, the 
[Z$_{{\rm Mg}b}$/Z$_{\rm \langle Fe \rangle}$] and
[Z$_{\rm CN_2}$/Z$_{\rm \langle Fe \rangle}$] values might help to put constraints on the assembly history of a galaxy.
In the case of NGC\,357, we find [Z$_{\rm Mgb}$/Z$_{\rm \langle Fe \rangle}$]=0.3-0.4 dex, 
which is the value expected for an E/S0 galaxy of $\sigma\sim$180 km\,s$^{-1}$
according to the well-known correlation between 
[Z$_{\rm Mg}$/Z$_{\rm \langle Fe \rangle}$] and $\sigma$ for early-type objects
\citep[e.g.,][]{1999MNRAS.306..607J,2000AJ....120..165T,2000MNRAS.315..184K,2004ApJ...601L..33V,2004ApJ...609L..45C}.
Note that an extrapolation
of the grid is necessary to estimate the [Mg/H] abundance, since the models do not extend beyond 
[Z/H]=0.2 dex. 
We also find a supersolar [Z$_{\rm CN_2}$/Z$_{\rm \langle Fe \rangle}$] ratio for the bulge of NGC\,357, 
that reaches $\sim$0.2 dex. 

Note that the bulge of NGC\,357 shows subsolar values for the Ca\,4227 index. 
This is the expected behaviour for galaxies of similar central velocity dispersion. 
In fact, rather than resembling other $\alpha$ elements as the Mg, Ca tends to get subsolar values
\citep{1997ApJS..111..203V}.

In order to find out if the formation of the bulge of NGC\,357 has followed
a distinguished process due to the presence of the inner bar, we compare our results
with those obtained for various elliptical galaxies and other spiral bulges.
Thus, we have overplotted the index values for two Virgo ellipticals
with a central $\sigma\sim$180 km\,s$^{-1}$, taken 
from the sample of \citet{2006ApJ...637..200Y}. These measurements appear
as two open circles in Figure \ref{fig:spBB}. 
Moreover, we have selected a subsample of four early-type barred galaxies from the works by
\citet{2007A&A...465L...9P,2009A&A...495..775P} and \citet{2011A&A...529A..64P} 
with central velocity dispersions similar to that of NGC\,357.
Two galaxies host a single bar (NGC\,1169 and NGC\,1358), and
two are double-barred systems (NGC\,2859 and NGC\,2962). For these objects, 
we have measured the mean line-strength indices inside the 
bulge region from the values at a resolution of $\sigma\sim$180 km\,s$^{-1}$.
The results, when possible, are also shown in Figure \ref{fig:spBB} 
as grey filled circles. In general, the age, metallicity and $\alpha$-enhancement
obtained for NGC\,357 resemble very closey those of the 
comparison bulges. The age and metallicity of the Virgo ellipticals are also in agreement
with NGC\,357, although they show a larger [Z$_{\rm CN_2}$/Z$_{\rm \langle Fe \rangle}$]
enhancement, with a mean value above $\sim$0.3 dex. 

The results for NGC\,357 from the line-strength analysis are also verified by the cross-correlation method.
Figure \ref{fig:ccBB} shows the 
peak heights resulting from cross-correlating the bulge spectra with SSPs of different ages and metallicities, 
following the methodology described in Section \ref{sec:cc}. From both spectral ranges, we
obtain a supersolar value ([Z/H]$\sim$0.2 dex) for the mean luminosity-weighted metallicity, 
and a final age of 7-8 Gyr (although the error regions span a wider range). So far,
the cross-correlation method alone is not able to clearly distinguish a mean luminosity-weighted
stellar population above $\sim$5 Gyr, due to the flattening of the cross-correlation
peak heights versus age profiles and the amplitude of the error bars.
For younger ages, Figure \ref{fig:ccBB} shows that the curves are steeper, allowing to
disentangle between different populations very efficiently. However, the fact that 
the results obtained from this procedure directly agree with the line-strength analysis,
points out the potential power of this method. Note that the 
age-metallicity degeneracy is evident in Figure \ref{fig:ccBB}: the shoulder
of each metallicity curve (i.e., the point where there is a clear transition in the slope)
occurs at older ages for lower metallicities.

\subsubsection{Stellar populations of the inner and outer bars}\label{sec:gradients}

The relative ages and metallicities between bulge, inner bar and outer bar of NGC\,357 are
shown in Figure \ref{fig:spALL}, where the age indicators H$\gamma_\sigma$ and
H$\delta_F$ are plotted against various metallicity indicators 
(including the new Fe4383$^{SR}$ definition) and the \emph{G}-band. In this
case we are limited to indices in the blue range, since it is the only spectral range
available for the two bars.

In the first panel of Figure \ref{fig:spALL} we plot H$\gamma_\sigma$  
against Fe4383$^{SR}$, which provide the most orthogonal grid
thanks to the age-disentangling power of H$\gamma_\sigma$.
We find that the bulge and inner bar show approximately
the same mean luminosity-weighted age ($\sim$8 Gyr) and solar metallicity. 
The outer bar also presents a similar age as the inner structures,
but it tends to be less metal-rich ([Fe/H]$\sim$-0.4 dex). However, the error bars are large due
to the S/N requirements of the H$\gamma_\sigma$ index. For this reason it is necessary
to complement these results with those from the less orthogonal H$\delta_F$ 
versus Fe4383$^{SR}$ grid, which confirms the previous estimations for the bulge and inner bar.
The trend towards lower metallicites for the outer bar is also verified. 
In fact, the only difference we find between the two plots
is that the outer bar seems to be slightly older when looking at 
the H$\delta_F$ versus Fe4383$^{SR}$ grid. However, taking into account the error bars and
the results from the rest of the plots (particularly 
H$\delta_F$ versus CN$_2$ and H$\gamma_\sigma$ versus Ca\,4227), we
must conclude that the three structural components of NGC\,357 are nearly coeval.

Since we have measurements for the Fe4383$^{SR}$ and CN$_2$ indices, we can use the 
[Z$_{\rm CN_2}$/Z$_{\rm Fe4383^{SR}}$] ratio as a proxy for the [CN/Fe] abundance,
as we did in Section \ref{sec:bulge} with [Z$_{\rm CN_2}$/Z$_{\rm \langle Fe \rangle}$].
We find that, whereas the bulge and inner bar present an overabundance of
[Z$_{\rm CN_2}$/Z$_{\rm Fe4383^{SR}}$]$\textgreater$0.2 dex, the outer bar has a more enhanced
value above 0.4 dex.

To make a comparison with the age and metallicity of other main bars, 
we take into account the same galaxies we adopted as a reference in 
Figure \ref{fig:spBB}. We compute the mean values of the line-strength indices along 
the main bar at a resolution of $\sigma\sim$180 km\,s$^{-1}$ from the data of \citet{2009A&A...495..775P}.
These comparison values are shown in Figure \ref{fig:spALL} and the plotted errors take into 
account the variation of the line-strength indices along the main bar. Like for the bulge
case, the outer bar of NGC\,357 shows no particular features when compared to the other 
main bars. Indeed, its age and metallicity are within the ranges defined by the control sample.

We apply the cross-correlation method only to the inner bar spectrum since that of the outer bar
has too low S/N. The mean luminosity-weighted 
age and metallicity are consistent with those derived from the analysis of the line-strength indices. 
Figure \ref{fig:ccIB} shows that
although the maximum of the cross-correlation peak heights is obtained for $\sim$4.5 Gyr, 
the solution spans over a wider range (from about 4 to 10 Gyr) when taking into account 
the error regions. The resulting metallicity, despite the partial overlapping between the solar 
and supersolar curves, is [Z/H]$\sim$0.2 dex.

\section{Discussion}\label{sec:dis}

The various pieces of evidence provided by this work lead
to two main questions: which is the real structure of NGC\,357, and how this galaxy has been formed.
The analysis of the photometry and kinematic constrains the structural composition
of NGC\,357, which reveals itself as a really complex galaxy. 
On the other hand, the study of the stellar populations provides 
surprisingly homogeneus results and sheds light on the most probable assembly process
for our galaxy. For the sake of clarity, we split the Discussion in two Sections, devoted
to each of these questions.

\subsection{Structural components of NGC\,357}

Photometrical studies have confirmed that NGC\,357 is composed of, at least, 
four structures, namely bulge, inner bar, outer bar and disc 
\citep{2004A&A...415..941E,2005A&A...434..109A}.
The presence of the inner bar is also supported by the photometrical and
kinematical analysis performed in this work. In fact, the ellipticity and PA profiles
presented in Figure \ref{fig:mor}, and the velocity and velocity dispersion profiles
along the inner bar direction presented in Figure \ref{fig:kin}, show clear signatures 
of this small, secondary bar, as explained in Sections \ref{sec:mor} and
\ref{sec:kin}. 

Interestingly, the analysis of the kinematics also shows evidence of an 
additional, inner component: a kinematically decoupled structure appears at the centre
of NGC\,357, rotating faster than its surroundings. This decoupling spans approximately $\pm$2 arcsec
and matches in size with the observed central $\sigma$-drop.
The different structures that shape NGC\,357 have to account for all the 
signatures present in the kinematical profiles.
In this Section we 
state the two possibilities that match with the results obtained in this work.

\subsubsection{An inner disc and a  classical bulge}

The presence of a central velocity dispersion minimum
in a spiral galaxy was first reported by
\citet{1989A&A...221..236B}. Since then, the
number of galaxies showing a $\sigma$-drop has increased significantly. In fact,
\citet{2006MNRAS.369..529F} find this signature in at least
46\% of a sample of 48 early-type spirals, although other cases have been 
noticed in smaller samples \citep[e.g.,][]{2003A&A...409..459M},
particularly in barred galaxies and
even including late-type spirals \citep{2006MNRAS.367...46G}
and double-barred objects \citep{2001A&A...368...52E}.
The most accepted explanation for the formation of $\sigma$-drops
in disc galaxies is the star formation
at their central regions: the new stars acquire the kinematical properties
of the gas they are formed from; the dissipative nature of that gas is more efficient
in the central regions because of the higher density and converts the gas into a cold stellar component,
such as an inner disc,
with a lower velocity dispersion than the surroundings. This scenario is in
agreement with the N-body simulations of \citet{2003A&A...409..469W} and
\citet{2006MNRAS.369..853W}, who made use of bars to transport gas
to the central regions. 

The case of NGC\,357 seems to support
that hypothesis since the central decoupling in velocity might be due to an inner disc
whose low velocity dispersion would be the cause of the $\sigma$-drop.
Therefore, NGC\,357 is composed of at least five structures: bulge, inner disc, inner bar,
outer bar and disc.

The $\sigma$-drop means a local minimum
in the velocity dispersion with respect to the corresponding value for the bulge, which
reaches $\sigma\sim$180 km\,s$^{-1}$ or higher (the velocity dispersion profile should peak
at the centre, where the $\sigma$-drop masks the velocity dispersion of the bulge). 
This high velocity dispersion value for the bulge implies it is pressure supported
and classical. This result is supported by the work of \citet{2005A&A...434..109A}, who
find that the bulge of NGC\,357 follows the same fundamental plane as the ellipticals or 
classical bulges of other early-type objects. 

Within this context, the appearance
of the $\sigma$-hollows at the edges of the inner bar in the velocity dispersion profile 
is also well understood. The 
$\sigma$-hollows were first seen in a sample of
four double-barred galaxies analysed through integral-field spectroscopy
\citep{2008ApJ...684L..83D}.
By means of N-body simulations, they tested the possible scenarios
that might give rise to these hollows, showing that their size and amplitude
depend on two parameters: the relative 
contribution of the flux of the bulge to the total luminosity and 
the difference in the velocity dispersion of the two components.
Therefore, a galaxy hosting a classical bulge with a high velocity dispersion and a colder, inner
bar which dominates the total luminosity at its ends (where the light profile of the bulge
has already decayed) will clearly show the two $\sigma$-hollows at the bar edges, as for
the case of NGC\,357.
In this scenario, the bulge is the hottest structural component of NGC\,357 and the
velocity dispersion decreases for the outer structures.
Table \ref{tab:sigmas} indicate the velocity dispersion values for each component.
It is important to note that the actual velocity dispersion of the inner disc is
probably lower than the value given in Table \ref{tab:sigmas}, since 
the measured value includes the contribution of the bulge within the inner disc region.

There are two important drawbacks for this possible structural composition of NGC\,357. 
First, the $a_4>0$ signature of the inner disc should be detected in
the photometrical analysis presented in Section \ref{sec:mor}. This is the case
of the nuclear stellar discs observed in the centre of ellipticals and
bulges \citep[][and references therein]{2010A&A...518A..32M}. A careful
inspection to the inner $\pm$2 arcsec in Figure \ref{fig:mor} shows no evidence of
discy isophotes.
Second, the formation of the inner disc
is usually related to a recent star formation in the centre, but the analysis of the stellar populations
for NGC\,357 shows old ages for the central components; a significant young population such as that
of an inner disc should be noticed when studying the mean luminosity-weighted age, even if
it is mixed with an older, more massive component. Therefore, although this hypothesis seems promising,
it does not explain all the properties found in this work.

\begin{table*}
  \caption{Characteristic velocity dispersion values for the different structural components of NGC\,357
   within the context of the two possible scenarios. For the bulge and pseudobulge, these values represent
   their maximum velocity dispersion. For the outer bar and disc, the mean velocity dispersion values
   within the corresponding regions included in Figure \ref{fig:kin} are provided.}
  \label{tab:sigmas}
  \begin{tabular}{ccccccc}
    \hline
    \hline
                  & Scenario & (Pseudo)bulge & Inner disc & Inner bar & Outer bar & Disc\\
    \hline
    $\sigma$      & (1)      & $\geq$180     & $\sim$170  & $\sim$165 & $\sim$145 & $\sim$105 \\
    (km\,s$^{-1}$) & (2)      & $\sim$170     & -          & $\sim$180 & $\sim$145 & $\sim$105 \\\hline   
  \end{tabular}
\begin{flushleft}
\hspace{3cm}(1) Classical bulge + inner disc + inner bar + outer bar + disc;\\ 
\hspace{3cm}(2) Pseudobulge + inner bar + outer bar + disc.\\
\end{flushleft}
\end{table*}

%

\subsubsection{A pseudobulge}

The second possibility for the structure of NGC\,357 is that there is no inner disc.
If this is the case, the central kinematical decoupling and the $\sigma$-drop
have to be caused by the bulge itself, which therefore shows disc-like properties that indicate
it is a pseudobulge rather than a classical one. This hypothesis is strongly
supported by the S\'ersic index \citep{1968adga.book.....S} 
measured for the bulge of NGC\,357 by \citet{2005A&A...434..109A}; they 
performed a photometric decomposition of an \emph{I}-band image and
obtained \emph{n}=1.40$\pm$0.08,
which is compatible with a pseudobulge. 
Indeed, \citet{2008AJ....136..773F} study a sample of 77 galaxies and find 
that $\sim$90\% of pseudobulges have \emph{n}$\textless$2, whereas classical bulges
have \emph{n}$\textgreater$2.

Within this new context, the $\sigma$-drop is not actually a drop with respect to the
higher velocity dispersion of the bulge, but the maximum velocity dispersion of the pseudobulge itself.
Of course, the pseudobulge presents a not so high velocity dispersion, since it is rotationally
supported, and therefore it can be colder than the inner bar. In this case, the two peaks 
of the velocity dispersion values at either sides of the centre
have to be understood as signatures of the relatively hotter inner bar where this component dominates the total
luminosity. The $\sigma$-hollows appear where the bar ends and the light of the pseudobulge 
is the main contribution to the total flux. 
The surrounding structure, i.e., the outer bar, is also hotter than the outer parts of the
pseudobulge and for that
reason the velocity dispersion values increase again beyond the $\sigma$-hollows.
The characteristic velocity dispersions for each structure is indicated in Table \ref{tab:sigmas}.
This scenario is completely different from that of the galaxies shown in \citet{2008ApJ...684L..83D}.
However, 
the main conclusion on the origin of the $\sigma$-hollows 
remains as it is attributed to the contrast of velocity dispersions between components.

This hypothesis also explains the results from the analysis of the morphology of NGC\,357: the
ellipticity profile acquires approximately the same value at either sides of the peak due to the inner bar,
suggesting that the centre and the surrounding regions are dominated by the same structure. 
Concerning the dynamical support of the bulge, we estimate $V_{\rm
max}/\sigma\sim$0.23 and $\epsilon\textless$0.1 in the bulge region;
therefore, the bulge is consistent with an oblate spheroid flattened
by rotation \citep{2004ARA&A..42..603K}. Nevertheless, the
interpretration of the $V_{\rm max}/\sigma-\epsilon$ relation has to
be done carefully, since a complex kinematical structure of the galaxy
might conduct to misleading results, as it is shown in \citet{2011MNRAS.414..888E}.

\subsection{Formation of NGC\,357}

The analysis of the stellar populations of NGC\,357 is focused on three
single regions: bulge, inner bar and outer bar. The striking result obtained in
this work is that the bulge and inner bar show very similar stellar population 
properties, indicating that they were formed in the same process
or, at least, from the same source of stars. 
Moreover, the outer bar is also nearly coeval to the inner structures, although it 
presents a lower metallicity and a larger $\alpha$-enhancement. This result points
towards a faster assembly of the outer bar with respect to the bulge and inner bar.
Finally, we compare the bulge and outer bar of NGC\,357 with other bulges and main bars of 
single- and double-barred galaxies with similar central $\sigma$ values
\citep{2009A&A...495..775P,2011A&A...529A..64P}, finding no significant differences
among them.

The results obtained here for NGC\,357 indicate that theories 
claiming that inner bars play a major role in the secular evolution might not be suitable 
for this galaxy. If inner bars were
key to transport gas to the central regions and trigger star formation
there, a gradient in age and metallicity from outer to inner structures would be expected, 
presenting a younger population in the galaxy centre. 
Although our galaxy has gas content and probably some star formation 
is taking place in it, the stellar population analysis indicates that
this is not the main mechanism driving the evolution of NGC\,357.
Therefore, these results lead to the question on how NGC\,357 formed. 

Numerical simulations have predicted different scenarios for the formation
of nested bars systems, most of them requiring the presence of a dissipative component
that is dynamically disturbed and generates the secondary bar
\citep[e.g.,][]{1993A&A...277...27F,2001ApJ...553..661H,2002ApJ...565..921S,2004ApJ...617L.115E}.
Within this scenario, the inner bar should be younger than the outer bar.
However, the two bars of NGC\,357 have similar population properties, so they 
were formed at the same time or, most likely, the existing stars from the outer
bar or the disc redistributed and shaped the inner structures.
Therefore, only those numerical works that do not need gas to generate the double-barred
objects seem to be compatible with the observational evidences shown here for NGC\,357.
This is the case of the N-body simulations by \citet{2007ApJ...654L.127D}, who use
a disc and a rapidly rotating bulge (like a pseudobulge) 
to create a double-barred system. In these simulations, the inner bar
appears first, but this is probably a consequence of the initial presence of the pseudobulge: 
in the real Universe, the outer bar most likely is formed first,
so the pseudobulge would be generated from the gas flow along it.
This scenario reinforces the hypothesis of the pseudobulge for the case
of NGC\,357, since the bulge rotation is an essential condition for generating
the two bars within the \citet{2007ApJ...654L.127D} gas-free framework.

There is one possibility to reconcile the classical secular approach with the observational
results for NGC\,357, and this is to assume a very fast formation of the central components:
immediately after the outer bar was created, the gas flowed along it, triggered 
a quick star formation and formed the inner bar and the bulge. Within this scenario, the final ages of the 
three components would be very similar. The outer bar would be slightly older,
but this is consistent with our results taking into account the error bars. 
In this scenario a positive metallicity gradient for the gas along the outer bar is required.
Although this fact might be somewhat striking, from a stellar (not gaseous) point of view
there is evidence for such gradients. Indeed, \citet{2009A&A...495..775P} 
study a sample of 20 barred galaxies and find that some of them tend to become
more metal-rich at the ends of the main bars. Moreover, those objects have
older mean ages and higher central velocity dispersions than the rest of the sample, which
is in agreement with the properties found for NGC\,357.
Unfortunately, the S/N ratio of our data is not enough to measure gradients along
the main bar to check this possibility.

\section{Conclusions}\label{sec:con}

We have performed for the first time a detailed analysis of the morphological, 
kinematical, and stellar population properties
of a prototype double-barred early-type disc galaxy.
We put special emphasis on the inner bar in order to constrain its
role in the evolution of this galaxy.
The observational strategy consisted in taking very deep long-slit spectra along the directions
of the inner and outer bars and the major axis of the disc.

The presence of the two bars is clearly revealed by the ellipticity and PA profiles,
which show the usual sharp changes due to this kind of structure.
Moreover, the kinematical analysis also presents clear signatures due to
the inner bar: the double-hump profile in the velocity \citep{2001A&A...368...52E}
and the $\sigma$-hollows in the velocity dispersion \citep{2008ApJ...684L..83D}. 
These hollows are due to the contrast between the velocity dispersion values of the inner
bar and the central structure, that might be a classical bulge with a velocity dispersion
higher than the inner bar, or a pseudobulge with a lower velocity dispersion than the bar.
The relative contribution of the inner bar to the total flux also influences 
the size and depth of the $\sigma$-hollows.
It remains unclear if the $\sigma$-hollows or other related features may appear
for the case of the outer bar. Unfortunately, the S/N of the main bar spectrum is not sufficient
to explore the kinematics at its ends.

The analysis of the kinematics of NGC\,357 also reveals
a kinematically-decoupled structure rotating faster than its surroundings 
at the centre of the galaxy. This signature matches in size with a central $\sigma$-drop. 
NGC\,357 is the first
observed galaxy in which a $\sigma$-drop and the $\sigma$-hollows coexist. This fact
illustrates the differences between both signatures: whereas the former are just due to the presence of the
inner bar, the latter is caused by an inner, colder structure.
Due to the different nature of both
signatures, we want to stress that a double-barred galaxy might present
a $\sigma$-drop, the $\sigma$-hollows, or both, as in this case.

The structural composition of NGC\,357 has to account
for all the observational evidences found in the analysis of the photometry and kinematics.
The presence of the disc and the two bars is clear, but the central region turns
out to be more difficult to disentangle. In fact, two possible scenarios are
compatible with the results obtained in this work. The first possibility
is that NGC\,357 hosts a classical, hot bulge together with an inner disc, which
would be responsible of the decoupling in the velocity profile and the $\sigma$-drop.
The second scenario is that the bulge of NGC\,357 is, in fact, a cold pseudobulge, which
is directly related to the kinematical central signatures.

The main conclusion of this work is that the bulge and inner bar show similar stellar population
properties (age, metallicity and $\alpha$-enhancement), whereas the outer bar has no significant
difference in age but it is less metal-rich and more $\alpha$-enhanced, indicating
that it was assembled in shorter timescales than the inner structures. 
This result seems to discard for this galaxy the traditional
secular evolution scenario, in which the star formation triggered by the gas flown along the
outer structures is causing the formation of the central parts. Therefore, NGC\,357 has been
shaped by the redistribution of the existing stars, maybe in a \emph{secular} way in which stars
from the disc form the inner components, or maybe in an initial \emph{disc + pseudobulge} framework
so the inner bar is lately formed from the pseudobulge stars.
This last scenario is backed by the numerical work of \citet{2007ApJ...654L.127D}, who
create double-barred systems from a pure stellar disc and pseudobulge. In this scenario, the fast rotation
of the pseudobulge is the key element to generate bars without the gas contribution.

An analysis such as the one presented in this work for a 
larger sample of double-barred galaxies is required to derive general and robust
conclusions on the formation of these complex objects. Fortunately, the striking results 
obtained for NGC\,357 are motivating and shed light on the importance of carefully studying double-barred
galaxies, from the outer disc to the very central regions.

\section*{Acknowledgments}
We are indebted to Patricia S\'anchez-Bl\'azquez and Isabel P\'erez for providing
the line-strength measurements for the reference barred galaxies.
We are also very grateful to Jairo M\'endez-Abreu, Jes\'us Falc\'on-Barroso, Inma
Mart\'inez-Valpuesta, Nacho Trujillo, Reynier Peletier and Mina Koleva, whose help and useful comments
have been very important. 
Comments and suggestions from the anonymous referee have greatly improved this paper.
This work has been supported by the Programa Nacional de Astronom\'ia y Astrof\'isica of the Spanish 
Ministry of Science and Innovation under grant AYA2010-21322-C03-02.
EMC is supported by the University of Padua through grants CPDA089220,
60A02-1283/10, and 60A02-5052/11 and by the Italian Space Agency (ASI)
through grant ASI-INAF I/009/10/0.
This work benefits from observations made with the NASA/ESA Hubble Space Telescope, obtained from the
Hubble Legaxy Archive, which is a collaboration between the Space Telescope Science Institute
(STScI/NASA), the Space Telescope European Coordinating Facility (ST-ECF/ESA) and the Canadian Astronomy
Data Centre (CADC/NRC/CSA).

\appendix
\section{The newly defined Fe4383$^{SR}$ index}\label{sec:apFe}
\begin{figure}
\begin{center}
  \includegraphics[width=0.48\textwidth]{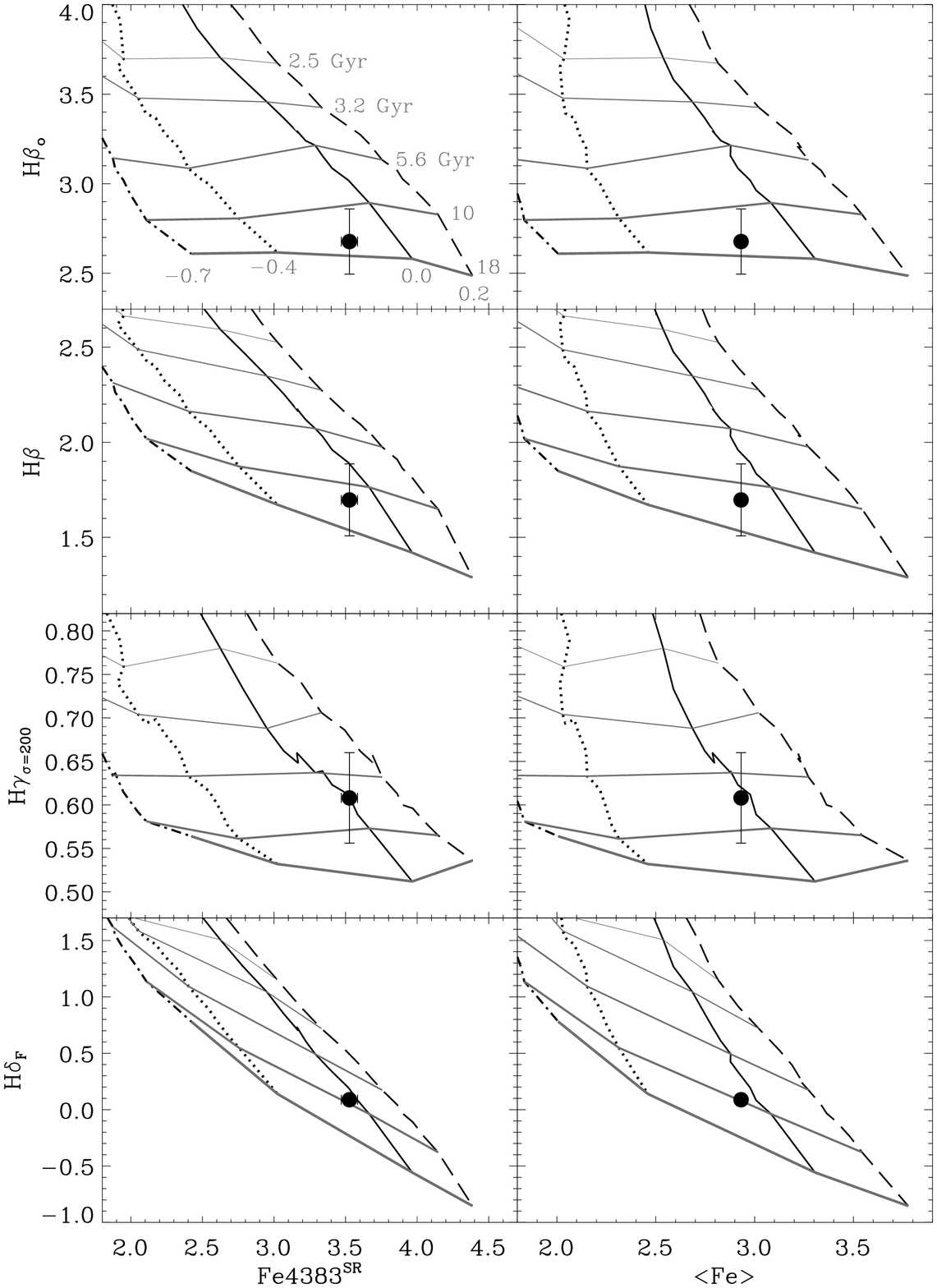}
  \caption{Comparison of the results obtained for the bulge of NGC\,357 using as metallicity
    indicators the newly defined
    Fe4383$^{SR}$ and the $\langle$Fe$\rangle$ indices. The model grids are the 
    same as in Figure \ref{fig:spBB}.}
  \label{fig:spFe4383}
\end{center}
\end{figure}

For the inner and outer bars of NGC\,357 we only have blue spectra, 
from 3990 to 4440 \AA. There is only one Iron index in the Lick system
within this range, at 4383 \AA. Unfortunately, the red pseudo-continuum of this
Lick Fe4383 index falls partially out of our blue spectra,
so we have redefined it in order to 
introduce a new Fe4383$^{SR}$ index suitable for our spectral range requirements.
Fe4383$^{SR}$ is named after \emph{Short Red} (due to its motivation) and keeps 
the blue pseudo-continuum of the Lick Fe4383 index, but modifies the range definitions 
for the main bandpass
and the red pseudo-continuum (see Table \ref{tab:fe}).
This new index has been tested with the bulge of NGC\,357, for which we have
data in two spectral ranges so we can measure additional Iron indices.
Figure \ref{fig:spFe4383} shows several age indicators plotted against Fe4383$^{SR}$ 
and $\langle$Fe$\rangle$. The corresponding measurements for the bulge of NGC\,357
are overplotted. The age and metallicity obtained from the two Iron index definitions 
are consistent. We therefore use Fe4383$^{SR}$ for estimating the metallicity 
in our blue spectra.

\section{S/N constraints for the cross-correlation method}\label{sec:apcc}
\begin{figure}
\begin{center}
  \includegraphics[width=0.48\textwidth]{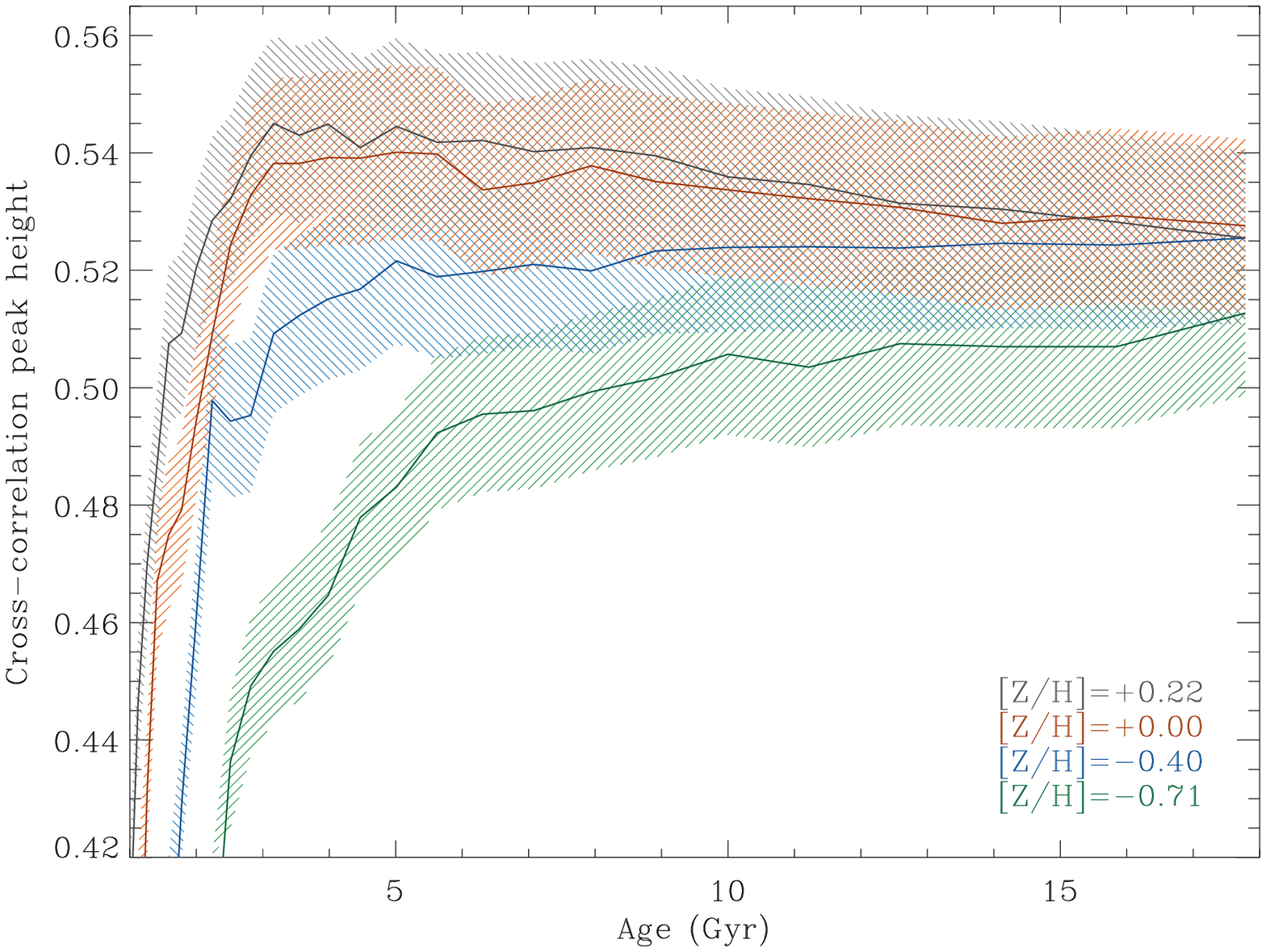}
  \caption{Same as Figure \ref{fig:ccBB} for a model of 8 Gyr and
    solar metallicity from the \citet{2010MNRAS.404.1639V} model library, once modified to 
    match the spectral resolution, range, dispersion and S/N ratio of the data for the outer bar
    of NGC\,357.}
  \label{fig:cctest}
\end{center}
\end{figure}

As explained in Section \ref{sec:cc}, we explored the requirements of the 
cross-correlation method through several tests with the \citet{2010MNRAS.404.1639V}
model library, by changing their spectral resolution, spectral range and S/N ratio.
We found that the S/N of the outer bar spectrum is not sufficient to disentangle its age
and metallicity. This fact is illustrated in Figure \ref{fig:cctest}, where we show
the results from aplying the cross-correlation method to a model of 8 Gyr and
solar metallicity, once cut and smoothed to mimic the observed spectrum for the outer bar. 
We also added gaussian noise to the model to match the S/N ratio of the data.
The error regions are so large that, for ages above $\sim$3 Gyr 
the results for the solar and supersolar metallicity are completely mixed.

\label{lastpage}

\end{document}